\documentclass[12pt]{iopart}

\usepackage{graphicx,cite}

\newcommand{\be}{\begin{equation}}
\newcommand{\ee}{\end{equation}}
\newcommand{\ba}{\begin{eqnarray}}
\newcommand{\ea}{\end{eqnarray}}
\def\C60{C$_{60}$}
\def\diff{\mathop{\rm\mathstrut d\!}\nolimits}

\begin{document}
\title[Phase coexistence in model protein solutions]
{Theory and simulation of short-range models
of globular protein solutions}

\author{G.~Pellicane, D.~Costa and C.~Caccamo\footnote[3]{Email address:
{\tt Carlo.Caccamo@unime.it}}}

\address{Istituto Nazionale per la Fisica della Materia (INFM) and\\
         Dipartimento di Fisica, Universit\`a di Messina \\
         Contrada Papardo, C.P. 50, 98166 Messina, Italy}

\begin{abstract}
We report theoretical and simulation studies of phase
coexistence in model globular protein solutions,
based on short-range, central, pair potential representations of the
interaction among macro-particles. After reviewing
our previous investigations of hard-core Yukawa and generalised
Lennard-Jones potentials,  we report more recent
results obtained within a DLVO-like description of
lysozyme solutions in water and added salt.
 We show that a one-parameter fit of this model based on
Static Light Scattering and Self-Interaction
Chromatography data in the dilute protein regime, yields
demixing and crystallization curves
in good agreement with experimental
protein-rich---protein-poor and solubility envelopes.
The dependence of cloud and solubility points
temperature of the model  on the ionic strength is also investigated.
Our findings highlight the minimal assumptions on the
properties of the microscopic interaction
sufficient for
a satisfactory reproduction
of the phase diagram topology of globular protein solutions.
\end{abstract}

\section{Introduction}

  Phase coexistence in aqueous solutions of globular proteins,
like for instance lysozyme in water and added salt,
has been intensively investigated in most recent years from 
experimental and theoretical
points of view (see e.g.~\cite{muschol95,zamora,belloni,muschol97%
,tardieu,zamora1,sic}
and references).
Such an interest is primarily intended to clarify
the mechanism of protein crystallization,
a process whose control,
though of crucial importance for the study of protein structure,
has not yet been characterised in terms of rigorous
and predictive experimental protocols~\cite{chayen,pherson}.

  Indeed, the literature documents a diffuse resort to
trial-and-error procedures
in the attempt to grow ``good'' crystals from the 
mother solutions~\cite{pherson};
this is also due to a lack of an accurate microscopic
description of the overall phase behaviour,
which depends on many factors:
for instance, even assuming that a satisfactory description
of the interactions among ``bare''  proteins were available,
understanding how these interactions are altered by
the solution variables is a difficult task, also considering that
specific salt effects determine a ranking of various ions,
which is
well-known as the Hofmeister series~\cite{broide}.
Despite the strong dependence on solution variables,
several globular proteins solutions
share important features, when
crystallization conditions are approached.
In particular, the second virial coefficient $B_2$
of many protein solutions
falls in a narrow window of small
negative values,
whenever the transition takes place~\cite{george}.
In close relationship with such an occurrence,
 the phase diagram is characterised by a metastable
protein-protein demixing region located
just below the solubility line~\cite{muschol97},
a peculiar property
of crucial interest,
since experimental and numerical studies
have shown that the proximity of the critical point to the solubility
line should favour the formation
of good crystals~\cite{muschol97,tenwolde,haas2}.
Besides the crystallization properties, the
protein demixing binodal plays an
important role also in several human diseases~\cite{eaton}.

These experimental evidences have been
rationalized in terms of effective
protein-protein interactions described through short-range,
centrosymmetric potentials~\cite{zamora,lomakin,belloni,%
piazza2,poon,zamora1},
and such models have
been intensively investigated by means of simulation
and liquid state theories.
On the other hand, the rich phase behaviour
of short-range attractive models has challenged
a straightforward generalisation to such systems
of concepts and methods derived from the theory of simple fluids.
In fact, it has been recognised (see~\cite{louis} and references)
that the onset of
freezing, and generally
the overall phase behaviour, is substantially affected
 by the ``perturbative'' part (with respect to the repulsive core)
of the potential, rather than being dominated  by excluded-volume and packing
(i.e. by entropic) effects, as is the case in the {\it van der Waals
picture} of simple liquids~\cite{chandler}.
The observation of two glassy states, originating
from different cage mechanisms which drive the structural arrest
of the system (see~\cite{dawson,foffi} and references),
recently confirmed by experiments~\cite{chen},
is another aspect of the unusual physical properties of systems
interacting through short-range forces.

In this context, one of the most extensively studied model is
represented by the hard-core Yukawa fluid (HCYF).
In fact, early simulations~\cite{hagen} for this system
showed that, when the decay of the attractive tail is sufficiently
fast, the phase diagram is topologically similar to that of most
crystallizing protein solutions, i.e. with a metastable
liquid-vapour binodal lying beneath the sublimation line.
 Yukawa fluids have been studied by other authors (see~\cite{report}
and references)
and by us~\cite{noi1,noi2,noi3}, in terms of integral
 equation theories (IET) of liquids~\cite{hansen}.
The utility of theoretical approaches, along with computer
simulation investigations, arises from several circumstances. In
fact, simulations of systems characterised by short-range
attractive forces, as in the case of charge-stabilised protein
solutions, might be affected by ergodicity problems; similarly,
the extension to more realistic multi-component cases (where
proteins, solute ions and water molecules are explicitly
considered) faces with severe difficulties, related to the strong
size asymmetry of the particle species, and to the usually high
dilution of the macromolecules. Such studies would greatly benefit
of the availability of reliable theoretical approaches to deal,
for instance, with the multicomponent configuration of the system.
We have devoted a considerable effort~\cite{noi1} to test refined
IETs against rigorous simulation data, starting from the simplest
one component HCYF case. We succeeded in particular in the
optimisation of the thermodynamic consistency, a constraint which
is often imposed in the solution of IETs in order to improve their
accuracy~\cite{GC}. In this case, we have investigated both the
HCYF and a simple DLVO model~\cite{dlvo0} of proteins, widely used
in colloid physics. 

As for the determination of the solid-fluid coexistence, we have
calculated the freezing line of envisaged models~\cite{noi1} in
terms of one-phase freezing criteria proposed by various
authors~\cite{hans:69,deltas}. We have assessed the accuracy of
such criteria in reference~\cite{simC60}, where theoretical
predictions for the Girifalco interaction potential of \C60 fullerene~\cite{girifalco}, 
characterised by a short-range decay, 
have been compared with the solid-fluid
equilibrium determined from free energy Monte Carlo simulations.
Interest to such calculations stems from the
observation~\cite{hagen} that the Boyle temperature of the \C60
model~\cite{girifalco} is reproducible with a somewhat
short-range HCYF potential. 
 Moreover, the Girifalco potential is
currently employed as a robust benchmark to test the performances
of new simulation strategies and theoretical approaches, when
applied to fluid with short-range 
interactions~(see~\cite{klein,fartaria,stell}). 
We summarise in section~2 our
most recent investigations of the HCYF and the Girifalco potentials.

We also review in section~2 our results~\cite{ballone} on the
crystallization processes occurring in another model for protein
solutions, namely the generalised Lennard-Jones potential early
proposed by ten Wolde and Frenkel~\cite{tenwolde}.
The phase diagram of the model was calculated
in reference~\cite{tenwolde} through computer simulation
estimates of the free energy, and the most favourable conditions
for crystal growth from the solution were identified.
We have carried out extended molecular dynamics calculations,
in order to compare our results for 
the crystallization kinetics with
the free energy calculations  reported in~\cite{tenwolde},  and
with experimental evidences in real protein solutions.

  In section~3 we report on recent
investigations~\cite{noi4a,noi4b,noi5} of the
phase diagram of lysozyme solutions in water and sodium chloride
added salt, modelled in terms of
a flexible representation
of protein-protein interactions,
where the forces acting among macromolecules are described by means of
a DLVO-like potential~\cite{dlvo0}.
We have been prompted to such an investigation
since other models~---~including the HCYF and
the generalised Lennard-Jones potentials~---~have hitherto
yielded only a qualitative reproduction of
the experimental phase diagram, and allowed at best to fit the
demixing line~\cite{lomakin,belloni};
these models also fail to capture
the sensitivity of the metastable demixing line
to solution conditions~\cite{zamora1}.
   We have performed both IET and simulation calculations for the
protein-rich---protein-poor coexistence curve
(the metastable  ``liquid-vapour''
binodal), and  determined the
solubility line through a simplified
version~\cite{asherie,sear,curtis,warren} of the
cell theory~\cite{devonshire} to estimate the
free energy of the solid phase.
We also show theoretical results for
the dependence of the cloud and solubility temperatures
on the ionic strength~\cite{noi4b},  in comparison
with experimental data for
lysozyme solutions.
Our conclusions follow in section~4.

\begin{figure}
\begin{center}
\includegraphics[width=7.0cm,angle=-90]{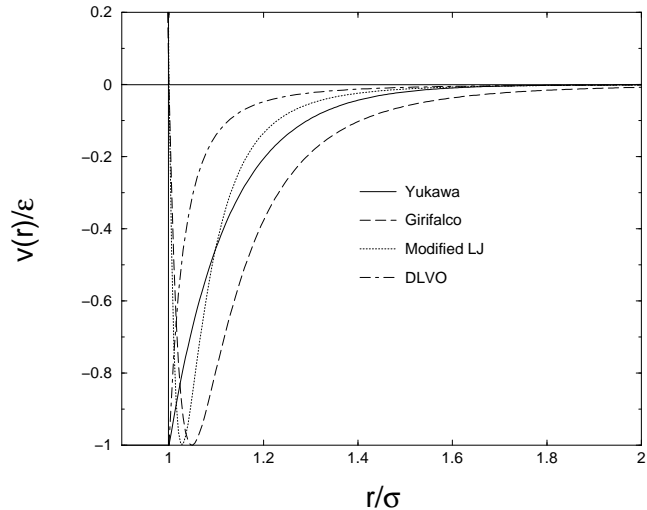}
\caption{\label{fig:pots} Model potentials presented in
this work. Full line: Yukawa potential with $z=7$, 
see equation~(\ref{eq:yukawa}); dashed line: Girifalco 
potential for \C60, equation~(\ref{eq:pot});
dotted line: modified Lennard-Jones potential
with $\alpha=50$, equation~(\ref{eq:vmlj}); 
dot-dashed line: DLVO model
with $\sigma = 3.6$\,nm, $Q=10$\,e, $A_{\rm H}=8.61k_{\rm B}T_{20}$
and $\delta=0.18$\,nm, see
equations~(\ref{DLVOdh})-(\ref{vrdlvo}). In order to compare 
among different models,
all potentials
are drawn in their reduced units of length
and energy, so that $v(r/\sigma=1)=0$ 
for the Girifalco and the modified Lennard-Jones potentials, respectively, 
and the minimum of the well depth 
is $-1$.}
\end{center}
\end{figure}

\section{Theoretical and simulation studies
of model fluids
interacting through short-range forces}

In the last few years we have undertaken an extensive
investigation of the phase behaviour of the
HCYF~\cite{noi1,noi2,noi3}, and assessed at the same time the
performances of most advanced liquid state theories presently
available (both semi-analytic and numerically solvable). 
We recall that the hard-core Yukawa potential is written as:
\ba\label{eq:yukawa} 
v(r) \ =\cases{\infty    & $r < \sigma$ \cr -\sigma\epsilon
\exp [-z(r-\sigma)] / r & $r \ge \sigma$ \cr} \,, \ea 
where
$\sigma$ is the hard-core diameter, $\epsilon$ is the potential
depth at closest contact and $z$ is the inverse screening
length (see figure~\ref{fig:pots}).
The properties of the HCYF are calculated
in the context of several liquid state theories, like for instance
the modified hypernetted chain 
approximation (MHNC,~\cite{MHNC}), the 
generalised mean spherical approximation (GMSA,~\cite{GMSA}) and the
self consistent
Ornstein-Zernike approximation (SCOZA,~\cite{pini}).
We shortly recall the basic features of such theories 
in the Appendix, and refer to other papers 
(see~\cite{report,MHNC,GMSA,pini} and references) for a detailed
presentation of their solution schemes.

The theoretical
results reviewed here concern an interaction range of the HCYF
corresponding to a realistic model of colloidal suspensions and
protein solutions~\cite{hagen,zamora,belloni}. We compare our
predictions with computer simulations carried out by other
authors~\cite{pini,lomba,hagen} or by ourselves.
Our investigations~\cite{noi1} show that the MHNC energies are
quite accurate for $z$ ranging from four to nine; the GMSA energies are
reasonably good at $z=4$ and show a $\sim10\%$ maximum discrepancy
from Monte Carlo (MC) data at the highest~$z$. The MHNC and GMSA
equations of state and compressibility are also within few
percents of the simulation results, at not too low temperatures.
The SCOZA results for the equation of state are quantitatively
accurate in any case. At lower $z$'s all theories become almost
quantitative.

\begin{figure}
\hbox to\hsize{%
\includegraphics[width=7.0cm]{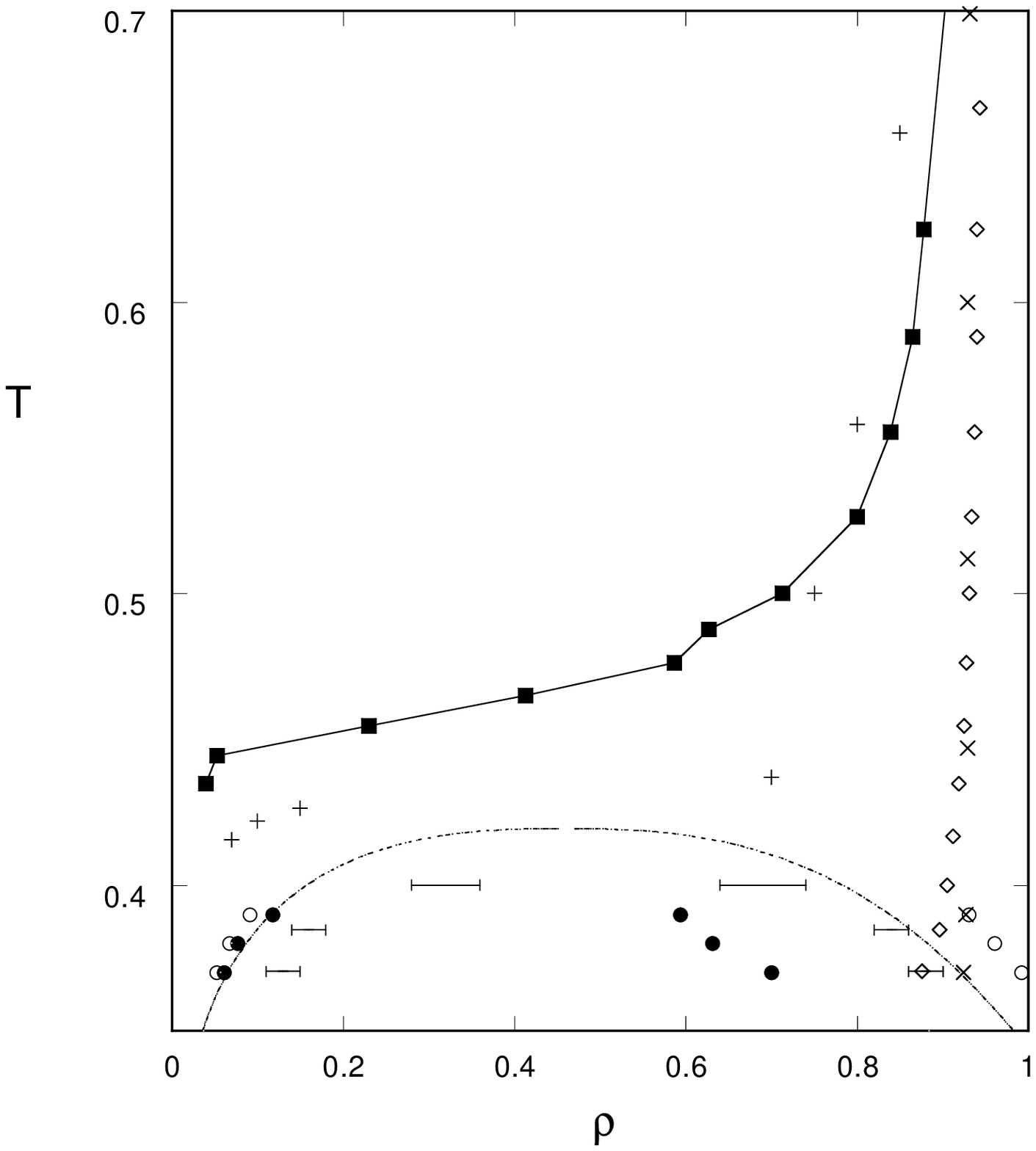}
\includegraphics[width=7.3cm]{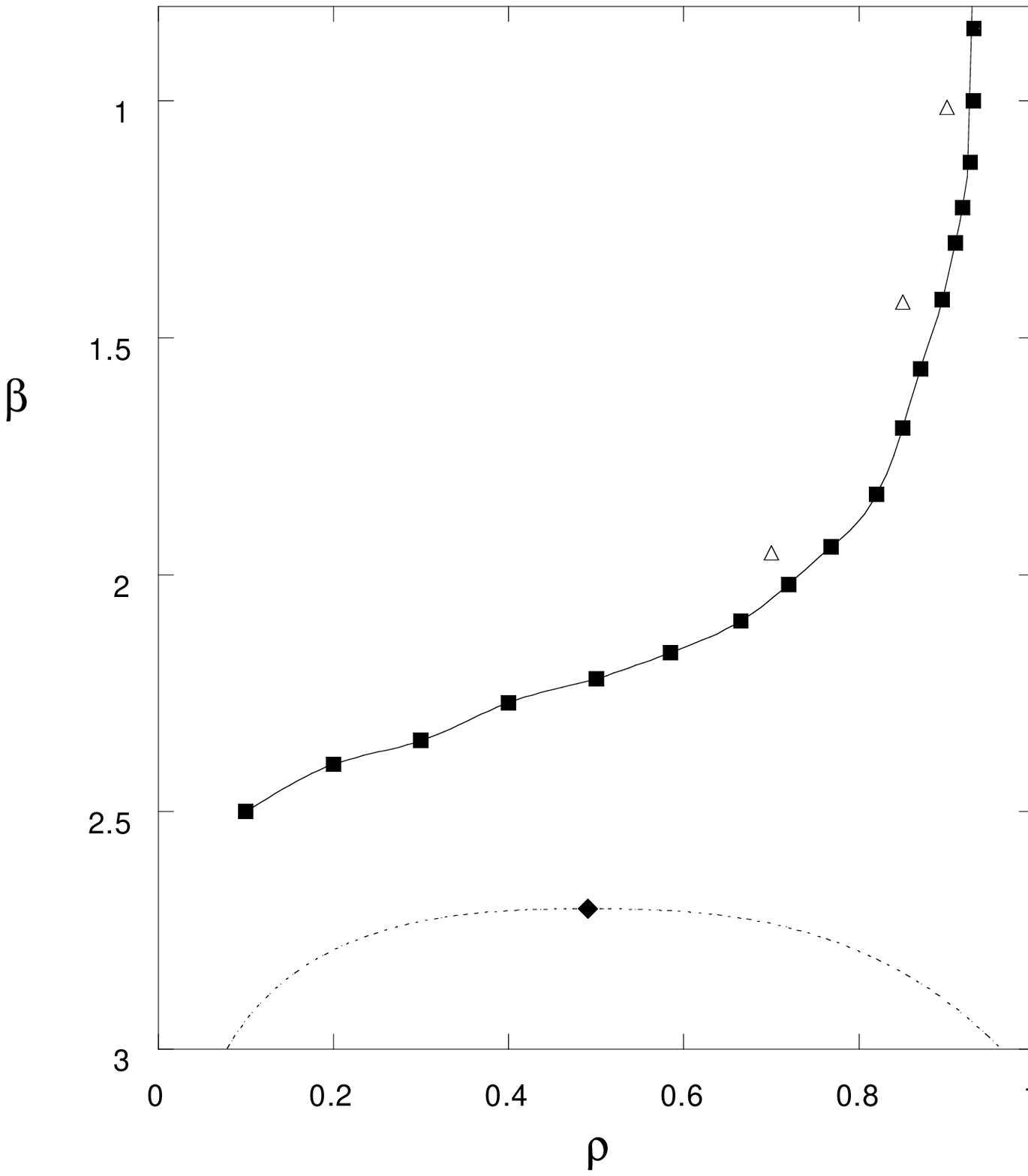}}
\caption{\label{fig:hcyf} Phase diagram of the HCYF;
temperatures are reported in units of $\epsilon/k_{\rm B}$,
densities in units of $\sigma^{-3}$, 
$\beta=\epsilon/k_{\rm B}T$. 
Left: $z=7$. Liquid-gas coexistence: horizontal bars, 
simulation results~\cite{hagen}; 
dotted line, SCOZA; full circles, GMSA;
open circles, MHNC. $\Delta s$=0 locus: diamonds, SCOZA;
pluses, GMSA; crosses MHNC; line with squares, sublimation
line of reference~\cite{hagen}.  Right: $z=9$.  
Triangles: GMSA predictions;
dotted line with diamond, SCOZA binodal and corresponding critical 
point~\cite{piniun};
line with squares: simulation results~\cite{hagen}. See
reference~\cite{noi1} for details.}
\end{figure}

Results for the phase diagrams are reproduced in
figure~\ref{fig:hcyf}, in the temperature-density ($T$-$\rho$)
representation, with the freezing lines estimated through
the one-phase criterion of reference~\cite{deltas},
amounting to the fulfilment of the condition 
that the so-called residual
multi-particle entropy $\Delta s$
vanishes at the phase transition.
 The limited applicability of
such structural indicators in the context of ``energetic''
fluids~\cite{louis} has been carefully discussed in
reference~\cite{simC60}. 
As visible in figure~\ref{fig:hcyf}, at
$z=7$ the GMSA qualitatively reproduces the relative location of
the binodal and sublimation lines
estimated in simulations
of reference~\cite{hagen}. SCOZA
results look fairly good; such an accuracy seems useful for the
investigation of the $z=9$ binodal, where no simulation results
are available. Also in this case the SCOZA binodal~\cite{piniun}
lies beneath the GMSA sublimation line; the latter one in turn,
fairly reproduces the high density portion of the sublimation
line. It appears that the combination of GMSA and SCOZA may
provide a description of the relative position of phase
coexistence lines of the HCYF in the very short-range interaction 
regime.

\begin{figure}
\hbox to\hsize{%
\hbox{\includegraphics[width=7.5cm]{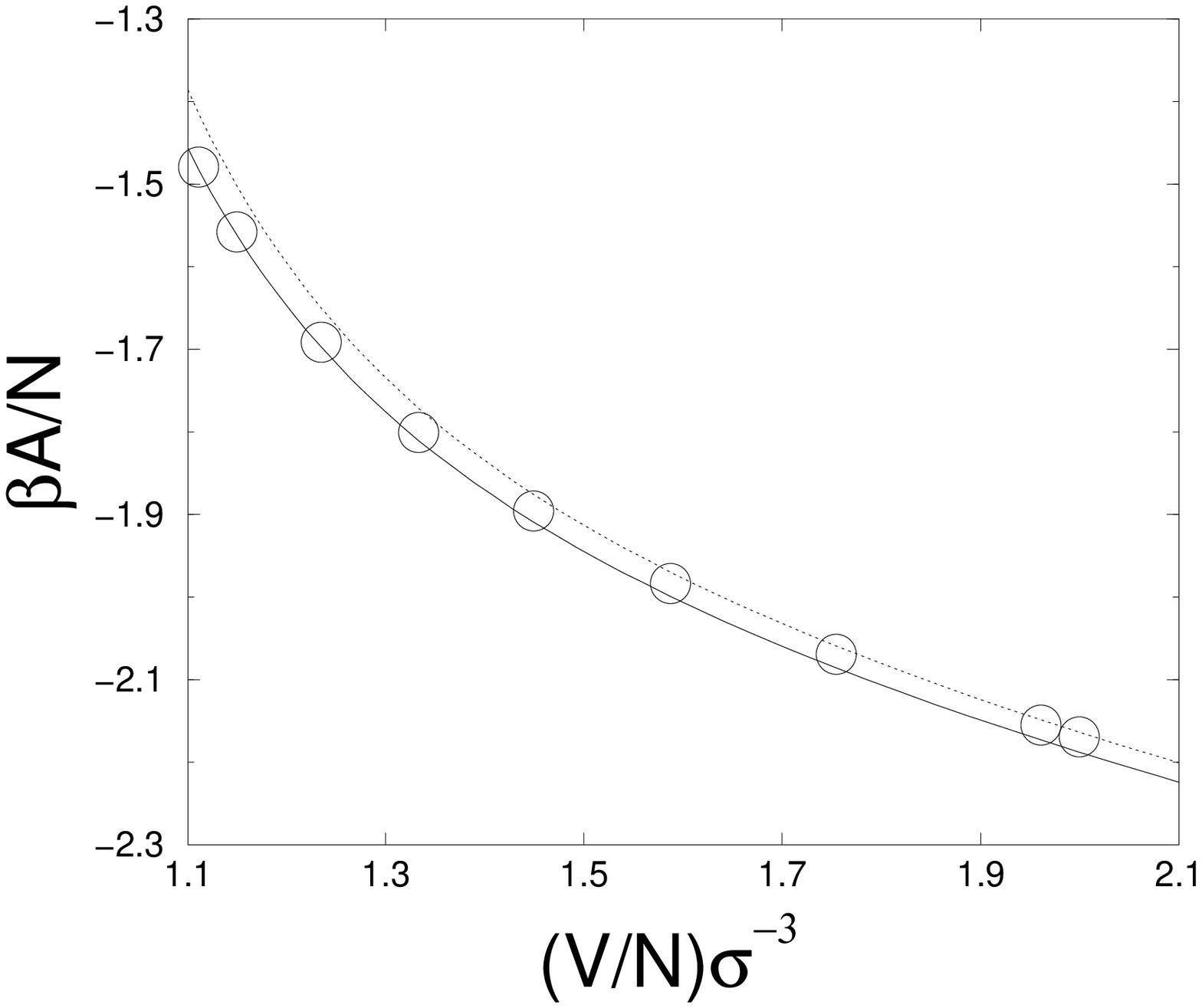}}
\hfill\hbox{\includegraphics[width=7.5cm]{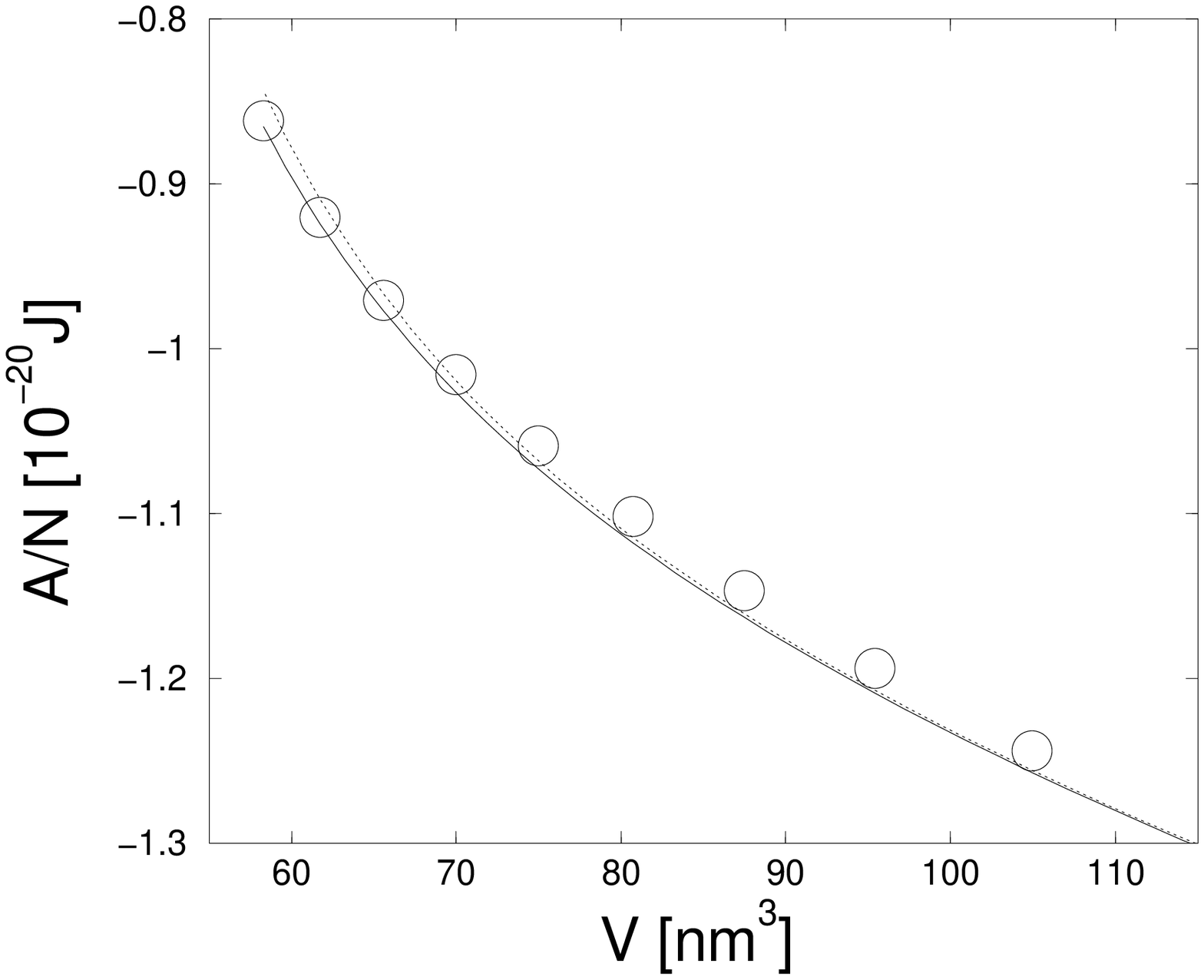}}}
\caption{\label{fig:hcyfdlvo} Helmholtz free energy as a function
of the system volume along supercritical isotherms for the HCYF
with $z=7$ at $k_{\rm B}T/\epsilon=0.5$ (left), and a DLVO model
with $\sigma = 3.6$\,nm, $Q=10$\,e, $A_{\rm H}=8k_{\rm B}T_{20}$
and $\delta=0.144$\,nm [see
equations~(\ref{DLVOdh})-(\ref{vrdlvo}) for symbols] at
$T=442.5$\,K (right). Open circles: MC simulations; dotted line:
MHNC; full line: HMSA. See reference~\cite{GC} for details.}
\end{figure}

In a subsequent study~\cite{GC}, we have exploited the sensitivity
of integral equation approaches like the MHNC and the Hybrid Mean
Spherical Approximation (HMSA,~\cite{HMSA}, see the Appendix), 
to the procedure
adopted in order to achieve the (partial) thermodynamic
consistency of the theory. 
We have analysed in particular the application of MHNC and HMSA to
the one-component HCYF
and to a system interacting through a 
short-range DLVO potential~\cite{dlvo0}
(see figure~\ref{fig:pots} and the next section for details).
We have found
that the enforcement of a global consistency constraint,
i.e. a procedure in which one imposes the equality of the
equation of state, as calculated
along different routes from structure to thermodynamics,
 is in general more accurate than the requirement of a local
consistency, based  on the equality
of the isothermal compressibilities.
The improvement associated to the global consistency
becomes crucial in the short-range interaction regime;
in particular, as visible in figure~\ref{fig:hcyfdlvo},
the Helmholtz free energy and the chemical
potential are almost quantitatively predicted
in the global consistency approach~\cite{GC}.
These results may help in the selection of 
integral equation approaches appropriate to determine
in a confident manner the phase diagram of both
pure fluids and binary mixtures of model short-range systems.

As far as a theoretical characterisation of solid-fluid
equilibrium is concerned, we have recently investigated~\cite{C60}
the solid phase of a pair, short-range model introduced
by Girifalco~\cite{girifalco} to
describe particle interactions in C$_{60}$:
\ba\label{eq:pot} \fl v(r) = 
-\alpha_1 \left[ \frac{1}{s(s-1)^3} +
                   \frac{1}{s(s+1)^3} - \frac{2}{s^4} \right] + 
     \alpha_2    \left[ \frac{1}{s(s-1)^9} +
                   \frac{1}{s(s+1)^9} - \frac{2}{s^{10}} \right] \,,
\ea 
where $s=r/d$, $\alpha_1=N^2A/12d^6$, and
$\alpha_2=N^2B/90d^{12}$; $N$ and $d$ are the number of carbon
atoms and the diameter, respectively, of the fullerene particles,
$A=32\times10^{-60}$~erg\,cm$^6$ and
$B=55.77\times10^{-105}$~erg\,cm$^{12}$ are constants entering the
Lennard-Jones 12-6 potential through which two carbon sites on
different spherical molecules are assumed to interact. For \C60,
$d=0.71$~nm while the node of the potential~(\ref{eq:pot}), the
minimum, and its position, are $r_0\simeq 0.959$~nm,
$\varepsilon\simeq 0.444\times10^{-12}$~erg and $r_{\rm
min}=1.005$~nm, respectively (see figure~\ref{fig:pots}).
 The
phase diagram of the model has been recently characterised in
terms of MC calculations of the free energies of the fluid and
solid phases~\cite{simC60,hasegawa}. We have also determined
theoretically  the free
energy of the solid phase in terms of a standard perturbation
theory (PT,~\cite{hansen}, see the Appendix). 
As visible in figure~\ref{fig:C60}, 
such an approach yields reliable predictions for the coexistence
properties  and, more generally, for the whole solid phase behaviour.
Since the PT embodies a
first-order expansion
of the model's free energy around the free energy
of a reference hard-sphere crystal, our results
noticeably 
imply that the hard-sphere
properties dominate, to a large extent, the structure of the solid
phase; conversely, as discussed in~\cite{simC60}, the structure of
the fluid phase is markedly affected by energetic aspects, related
to the attractive part of the potential. Our results further
support the use of PT to characterise the solid phase of systems
interacting through short-range forces, already documented in
references~\cite{foffi,amokrane,nava}.

\begin{figure}
\begin{center}
\hbox to\hsize{%
\hbox{\includegraphics[width=6.2cm,angle=-90]{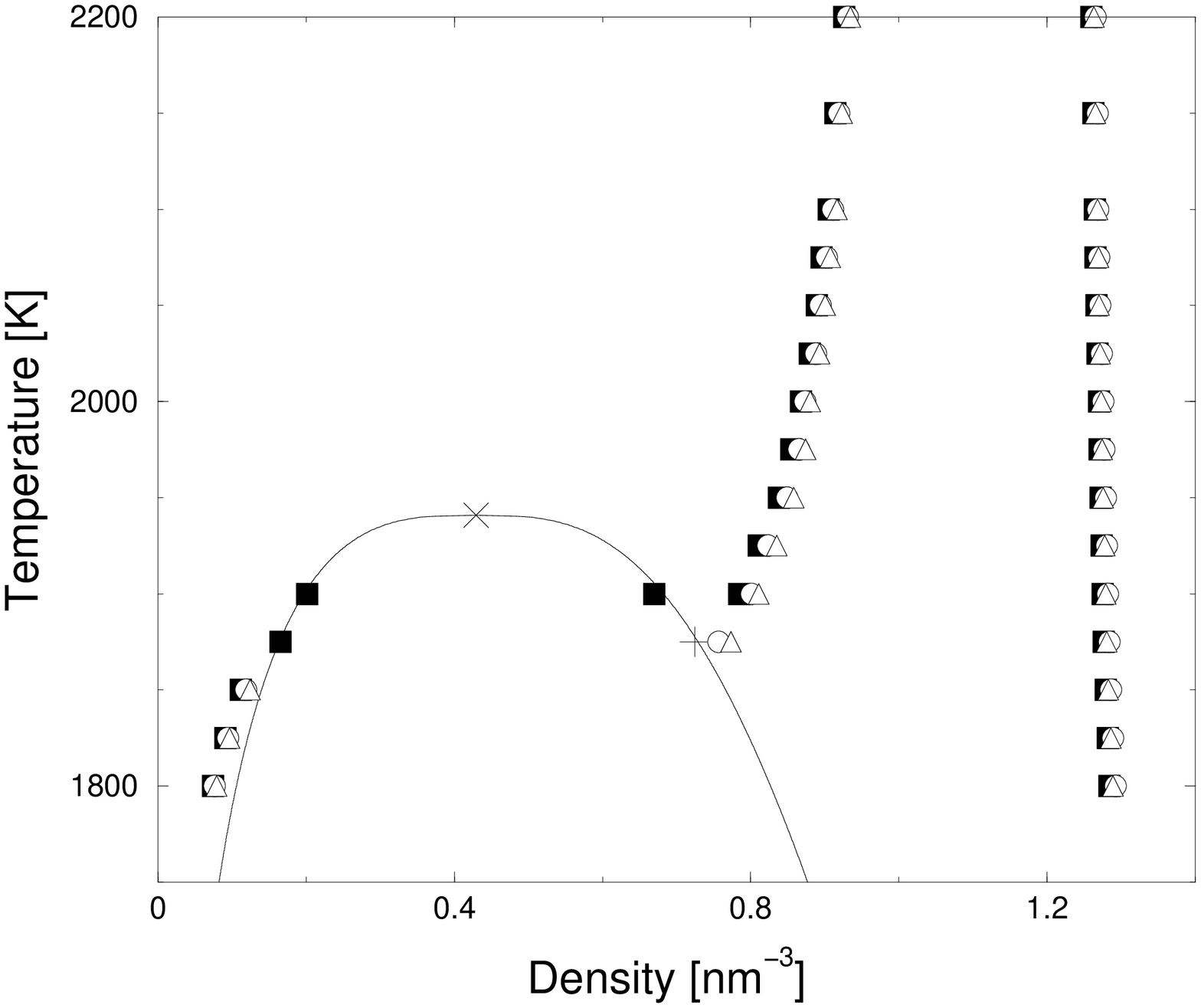}}
\hfill\hbox{\includegraphics[width=6.2cm,angle=-90]{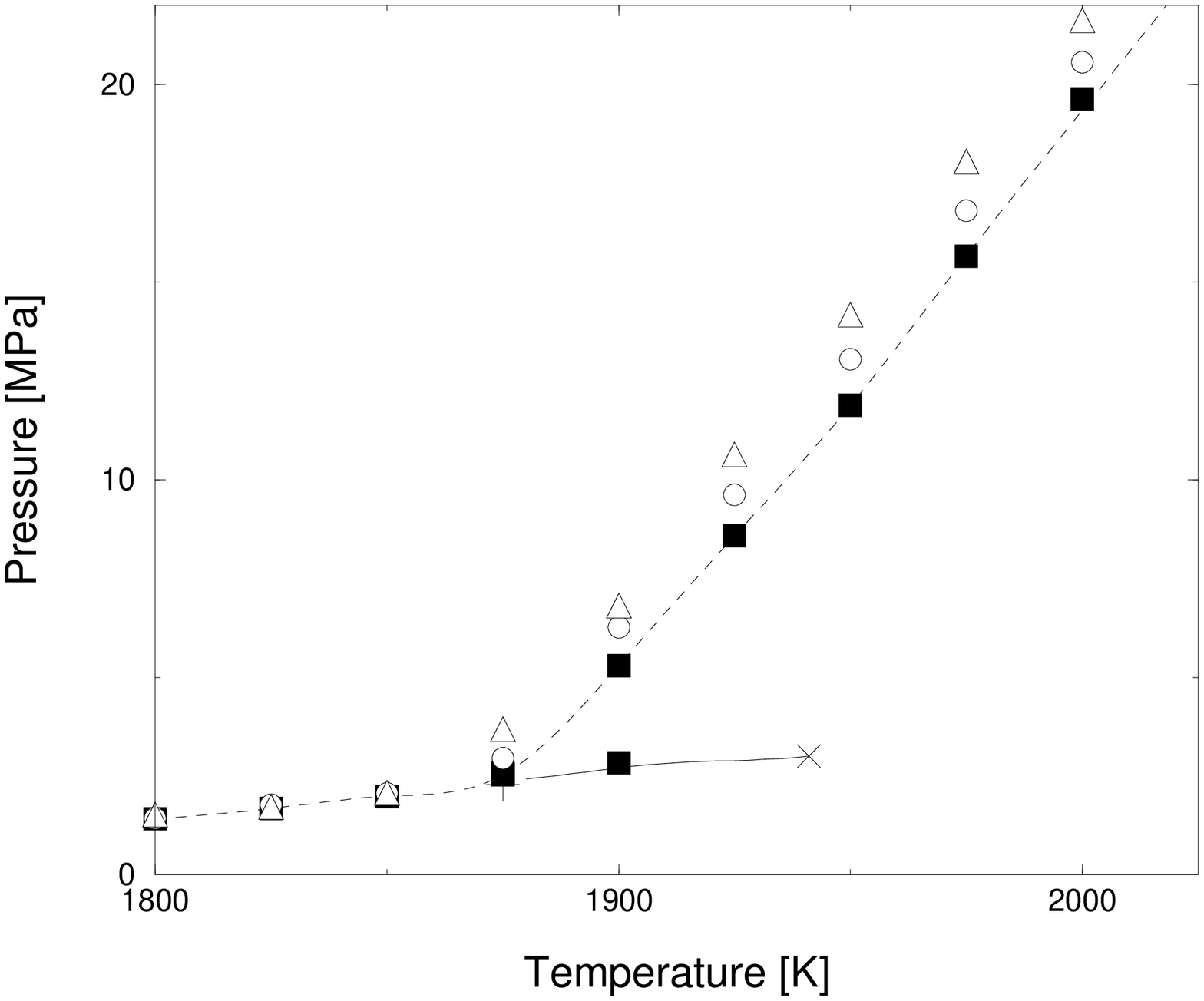}}}
\caption{\label{fig:C60}
Fluid-solid coexistence densities (left)
and pressures (right) of the Girifalco C$_{60}$ model.
PT predictions are used for the solid phase and MC
data for the fluid phase~\cite{simC60}.
Open symbols refer to different prescriptions for the
effective hard-core diameter entering the theory---see 
equation~(ref{eq:sigma1}) in the Appendix.
Squares are full Monte Carlo results~\protect\cite{simC60}.
The Gibbs Ensemble Monte Carlo liquid-vapour
coexistence line~\cite{Fucile} is also shown.
Dashed lines are guides to the eye.
See reference~\protect\cite{C60} for details.}
\end{center}
\end{figure}

\begin{figure}
\begin{center}
\includegraphics[height=20cm]{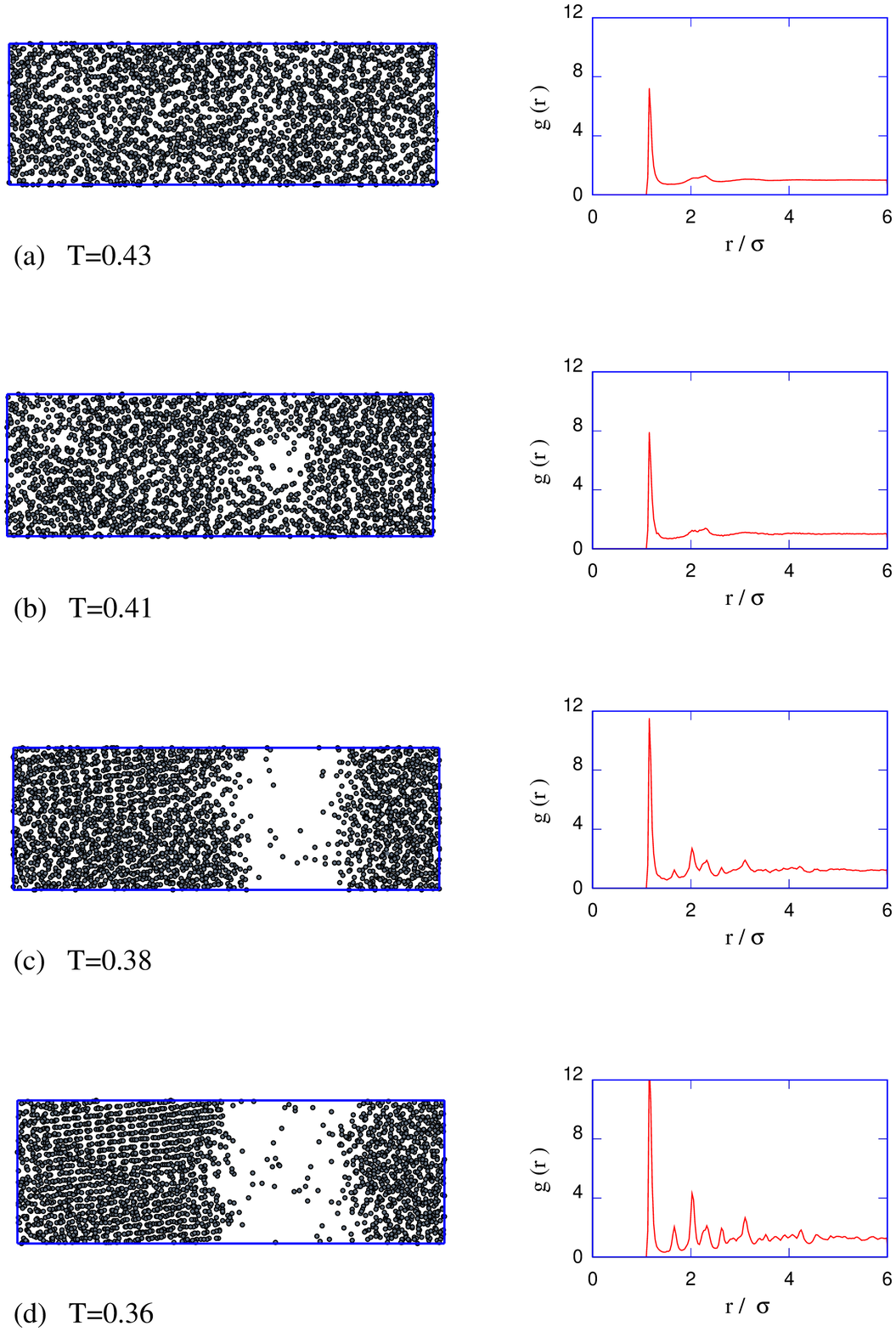}
\end{center}
\caption{\label{fig:t_down} Generalised Lennard-Jones 
potential---see equation~(\ref{eq:vmlj}).
Snapshots of MD configurations at progressively decreasing
temperatures (in units of $\epsilon/k_{\rm B}$),
together with the corresponding radial distribution
functions; (a) metastable homogeneous
fluid; (b) low-density bubble in a
high-density fluid, just below the metastable liquid-vapour
separation; (c) crystal nucleus in the
high-density fluid; (d) extended
defective crystal phase. See reference~\cite{ballone} for
details.}
\end{figure}

We have also investigated in a recent study~\cite{ballone}
the phase transformations
occurring in a system of spherical particles interacting through  a
short-range potential, obtained as
a generalisation of the Lennard-Jones interaction law~\cite{tenwolde}:
\be\label{eq:vmlj}
v(r)=\frac{4\epsilon}{\alpha^2} \left\{
        \left[ \left( \frac{r}{\sigma} \right)^2-1 \right]^{-6}
-\alpha \left[ \left( \frac{r}{\sigma} \right)^2-1 \right]^{-3}
\right\}
\ee
By changing the parameter $\alpha$, this expression gives rise to
a family of potentials whose shape ranges from a Lennard-Jones
like potential (for $\alpha \sim 1$) to potentials having a steep
repulsive part, followed by a deep and narrow attractive well,
approaching the sticky sphere limit with increasing values of
$\alpha$. In our computations, we set $\alpha=50$, thus recovering
the model early proposed in reference~\cite{tenwolde}
to study the phase behaviour
of globular protein solutions (see figure~\ref{fig:pots}).
We have focused on the
crystallization processes of this system~\cite{ballone}.
In particular, we have performed extensive molecular dynamics (MD)
simulations
of the metastability induced
by progressively cooling the fluid below the fluid-solid
and the (metastable) liquid-vapour coexistence at fixed density.
We have analysed in detail how the crystallization
process evolves in the simulated system, and the interplay of
crystallization with the metastable liquid-vapour separation.
In figure~\ref{fig:t_down} we show
a sequence of snapshots of MD simulations
at $\rho\sigma^3=0.5$
(i.e. a density moderately higher than the critical
density of the liquid-vapour binodal), for various decreasing
temperature in a path which crosses the binodal line
at a reduced temperature $T\simeq0.41$.
Model~(\ref{eq:vmlj}) for globular proteins is
admittedly a crude one. Nonetheless,
qualitative similarities
with the experimental behaviour still survive, as
for instance the enhancement of the crystallization
kinetics due to the presence of
the metastable liquid-vapour separation;
this phenomenon is very likely
determined by the presence of the
high density fluid, which decreases both the interfacial
free energy required to nucleate the crystal, and the amplitude of the density
fluctuations, necessary to reach the solid density.
Another interesting outcome is that the simulated
crystallization process
does not appear to be accompanied by any precursor
effect: in fact, when any of the criteria we devised to this aim was met,
the nucleation process  was in any case already well
underway. Our observations do not strictly prove, however, that precursors
do not exist, but only that the chain of events leading to the transition
involves the non-trivial coupling of several variables, as shown
experimentally in reference~\cite{schatzel}.

\section{Colloidal models for lysozyme solutions}

In a recent series of studies~\cite{noi4a,noi4b,noi5}
we have considered
a prototype globular protein solution, namely
lysozyme in water and
sodium chloride added salt, extensively characterised from the
experimental point of view
(see e.g.~\cite{muschol95,zamora,belloni,muschol97%
,tardieu,zamora1,sic}
and references).
In particular,
we have parameterised the physical properties of
the protein solution in the framework provided by the  DLVO
theory of charged colloidal suspensions~\cite{dlvo0}.
We recall that the
 DLVO potential of mean force is given by the sum
of a repulsive, Debye-H\"uckel contribution,
\be\label{DLVOdh} v_{\rm DH}(r) = \frac{1}{4\pi \epsilon_0
\epsilon_{\rm r}} \left[\frac{z_{\rm p}e} {1 + \chi_{\rm DH}
\sigma/2}\right]^2 \frac{\exp \left[- \chi_{\rm DH}(r -
\sigma)\right]}{r} \,, \ee
and an attractive part, represented by a short-range van der Waals
term:
\be\label{DLVOham} v_{\rm HA}(r) = -\frac{A_{\rm H}}{12} \left[
\frac{\sigma^2}{r^2} + \frac{\sigma^2}{r^2-\sigma^2} +
2\ln\frac{r^2-\sigma^2}{r^2} \right] \,. \ee
The resulting total interaction is:
\be\label{vrdlvo} v_{\rm DLVO}(r) \ =\cases{\infty& $r < \sigma +
\delta$ \cr v_{\rm HA}(r) + v_{\rm DH}(r)& $r \ge \sigma + \delta$
\cr} \,. \ee
Here $\sigma$ represents the effective diameter,
  $Q = z_{\rm p}e$ is the net charge (in electron units)
 of the macroparticle and
$A_{\rm H}$ is the Hamaker constant; $\epsilon_{\rm r}$ and
$\epsilon_0$ are, respectively, the (solution) relative and the
vacuum dielectric constants; $\chi_{\rm DH}$ is the inverse Debye
screening length. The Stern layer thickness $\delta$, related to
the intrinsic size of counterions which condense on the protein
surface, is introduced in equation~(\ref{vrdlvo}) to circumvent
the singularity of the attractive term at
$r=\sigma$~\cite{muschol95}.

\begin{figure}
\begin{center}
\includegraphics[width=7.0cm]{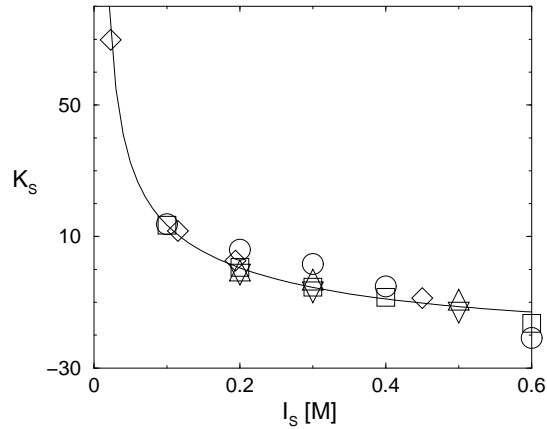}
\caption{Interaction factor as a function
of the solution ionic strength for lysozyme in water and NaCl added salt.
Circles: SIC~\cite{sic};
squares: SLS~\cite{zamora,zamora1};
diamonds: SLS~\cite{muschol95};
triangles up: SLS~\cite{tardieu} at
$T=303.15$\,K; triangles down: SLS~\cite{tardieu} at $T=293.15$\,K.
Full line: DLVO potential with $A_{\rm H}=8.61k_{\rm B}T_{20}$.
See reference~\cite{noi5} for details.}
\label{fig:lyso1}
\end{center}
\end{figure}

In our analysis,
we have fixed all parameters entering equation~(\ref{vrdlvo}),
but the Hamaker constant,
to values commonly accepted on the basis of several
experimental evidences for lysozyme
i.e. $\sigma=3.6$\,nm, $Q=10$\,e,
and $\delta=0.18$\,nm~\cite{muschol95,dlvoref,beretta}.
The Hamaker constant is used as a free parameter to fit
the experimental interaction factor defined as:
\be\label{DLVOnew}
K_{\rm S} = K_{\rm S}^{\rm HS} + 24 \int_{\delta/\sigma}^{\infty} x^2 \, \{1-\exp[-\beta
v_{\rm DLVO}(x)]\} \, {\rm d}x \,,
\ee
where $x=(r/\sigma -1)$ and $K_{\rm S}^{\rm HS}=8$
is the interaction
factor corresponding to a fluid of hard spheres.
The interaction factor is proportional to the second virial coefficient,
$B_2=K_{\rm S}\nu/2M$, where $\nu=0.703$\,ml/g and
$M=14300$\,Da are respectively the specific volume and the atomic mass of
lysozyme.
As shown in figure~\ref{fig:lyso1},
we best-fit Self Interaction Chromatography
(SIC) results, reported in reference~\cite{sic}, as well as
Static Light Scattering data from various sources
(SLS~\cite{muschol95,zamora,tardieu,zamora1}),
for a lysozyme solution in the range~$[0.1-0.6]$\,M NaCl
at a pH close to that of the experimental phase diagram~\cite{muschol97}.
Our fits are carried out in a temperature range
relatively far from the demixing region;
under these conditions the second virial coefficient of the
lysozyme solution is expected to vary moderately, and
almost linearly, with
the increase of the temperature.
An average Hamaker constant is then obtained,
$A_{\rm H}=8.61$
(in $k_{\rm B}T_{20}$ units) with a maximum half-dispersion equal to $0.26$.

The extension of a ``colloid physics approach'', as the DLVO
theory, to globular protein solutions has long been debated. In
fact, the literature reports either successful applications~(see
e.g.~\cite{poon,farnum,beretta,rowe}), or evidences of opposite
sign, concerning the accuracy and the theoretical foundation of
this picture~\cite{piazza,petsev,ninham}. Our observation is that,
in agreement with previous studies, the validation of the present
approach mainly derives from a positive, non-trivial,
parameterisation of phenomenological observations. In particular,
we envisage the functional form of the attractive potential in
equation~(\ref{DLVOham}) as flexible enough to accommodate not
only the van der Waals forces, but also other non-DLVO
interactions, as hydration effects~\cite{farnum,beretta,chernov}.
We assume in fact that these forces act to artificially inflate
the value of the Hamaker constant, which plays a different role
from the original one, only related to dipolar attractions. Within
this framework, specific effects, related for instance to the salt
identity, are taken into account only indirectly, through the fit
of experimental data. Although one could introduce additional
interaction terms (which can be salt-specific), in order to
increase the flexibility of the phenomenological potential, 
we shall show that 
a fine tuning of the
van der Waals amplitude
is sufficient to set
the proper order of magnitude of the effective protein-protein
interaction. In particular, this attractive term is the leading
contribution to the effective potential because repulsive
electrostatic interactions are almost screened out at biological
salt molarities, thus constituting a minor correction to the
potential of mean force.

\begin{figure}
\begin{center}
\includegraphics[width=8.0cm]{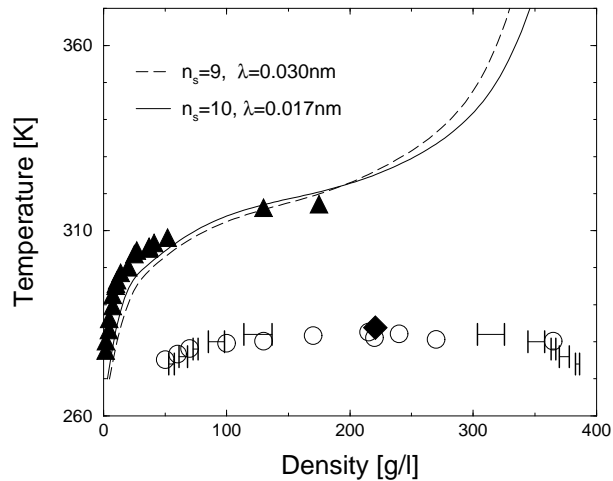}
\caption{Phase diagram of a lysozyme solution in water
and NaCl at  $I_{\rm s}=0.51$\,M.
Open circles and  full triangles:
cloud  and solubility points, respectively~\protect\cite{muschol97}.
Error bars with diamond:
DLVO demixing envelope and corresponding critical point.
Full and dashed lines: DLVO crystallization curves
with parameters for equation~(\ref{chemcrystal}) in the legend.}
\label{fig:lyso2}
\end{center}
\end{figure}

On the basis of the above model,
we have calculated the phase diagrams of the lysozyme
solution at $I_{\rm s}=0.51$\,M reported in figure~\ref{fig:lyso2}.
The protein-rich---protein-poor phase coexistence
line is determined by means
of Gibbs Ensemble Monte Carlo~\cite{GEMC}
simulations, carried out on a system composed of 1024 particles
(with a $3\sigma$ interaction cutoff).
As for the solubility line,
the free energy of the fluid phase is calculated
in the framework provided by
the refined HMSA theory~\cite{HMSA} whereas, for the solid phase,
we have used for the chemical potential
a simplified expression~\cite{asherie,sear,curtis,warren}
derived in the framework of the cell theory~\cite{devonshire},
\begin{equation}\label{chemcrystal}
\mu_{\rm CRY} = \mu_{\rm 0} -n_{\rm s}\epsilon_{\rm EFF}/2
-3k_{\rm B}T\ln(\lambda^*) \,.
\end{equation}
This equation provides a direct link with the essential properties
of the protein crystals, namely the average number of contacts,
$n_{\rm s}$, and the translational freedom along one axis,
$\lambda$ [with $\lambda^*=\lambda/(\sigma+\delta)$], of the
protein inside the unit cell, so that
$\lambda^3$ constitutes a rough estimate
of the volume accessible to the protein
inside the cage formed by its first neighbours; 
$\mu_0$ is the standard part of the
chemical potential, and $\epsilon_{\rm EFF}$ is the minimum of the
effective potential~(\ref{vrdlvo}). A constant value
$\log[V_p/(\sigma+\delta)^3]$~\cite{warren} has been included in
$\mu_0$, where $V_p$ is the protein volume. The coexisting fluid
branch is then found by equating the chemical potentials of both
phases~\cite{sear}. We fix the number of contacts between nine and ten,
in agreement with the estimates of reference~\cite{janin}; a
comparable number of contacts has been specifically reported for
lysozyme crystals in~\cite{NMR}. Our results for the phase diagram
are compared in figure~\ref{fig:lyso2} with the experimental
demixing and solubility lines reported in
reference~\cite{muschol97}. As visible, the theoretical demixing
envelope coincides with the experimental {\it locus} of cloud
points with a critical temperature $T_{\rm cr}\simeq 284$\,K, very
close to the experimental outcome $T_{\rm cr}=282.5$\,K. The
theoretical crystallization
 boundary, calculated with
$n_{\rm s}=9$ and $\lambda = 0.03$\,nm in equation~(\ref{chemcrystal}),
 is also in good agreement with the experimental curve.
We obtain similar results if another plausible
number of contacts, namely $n_{\rm s}=10$
(with $\lambda = 0.017$\,nm) is assumed (see figure~\ref{fig:lyso2}).
All values of $\lambda$ are compatible with the temperature factors
of the Bragg intensities in lysozyme crystals,
where the typical range for $\lambda$ is $[0.014-0.033]$\,nm~\cite{doucet}.

\begin{figure}[!t]
\begin{center}
\hbox to\hsize{%
\hbox{\includegraphics[width=6.0cm,angle=-90]{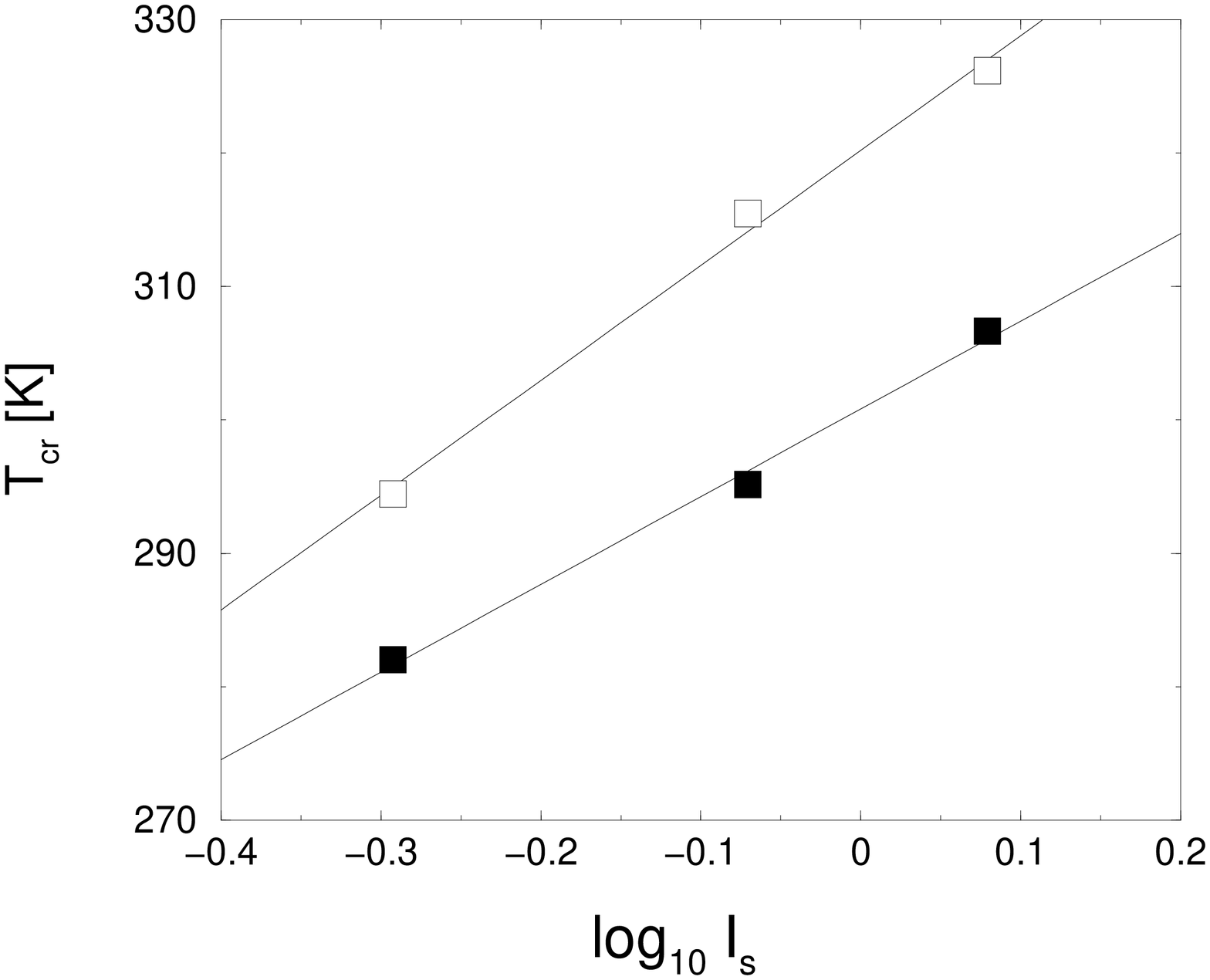}}
\hfill\hbox{\includegraphics[width=6.0cm,angle=-90]{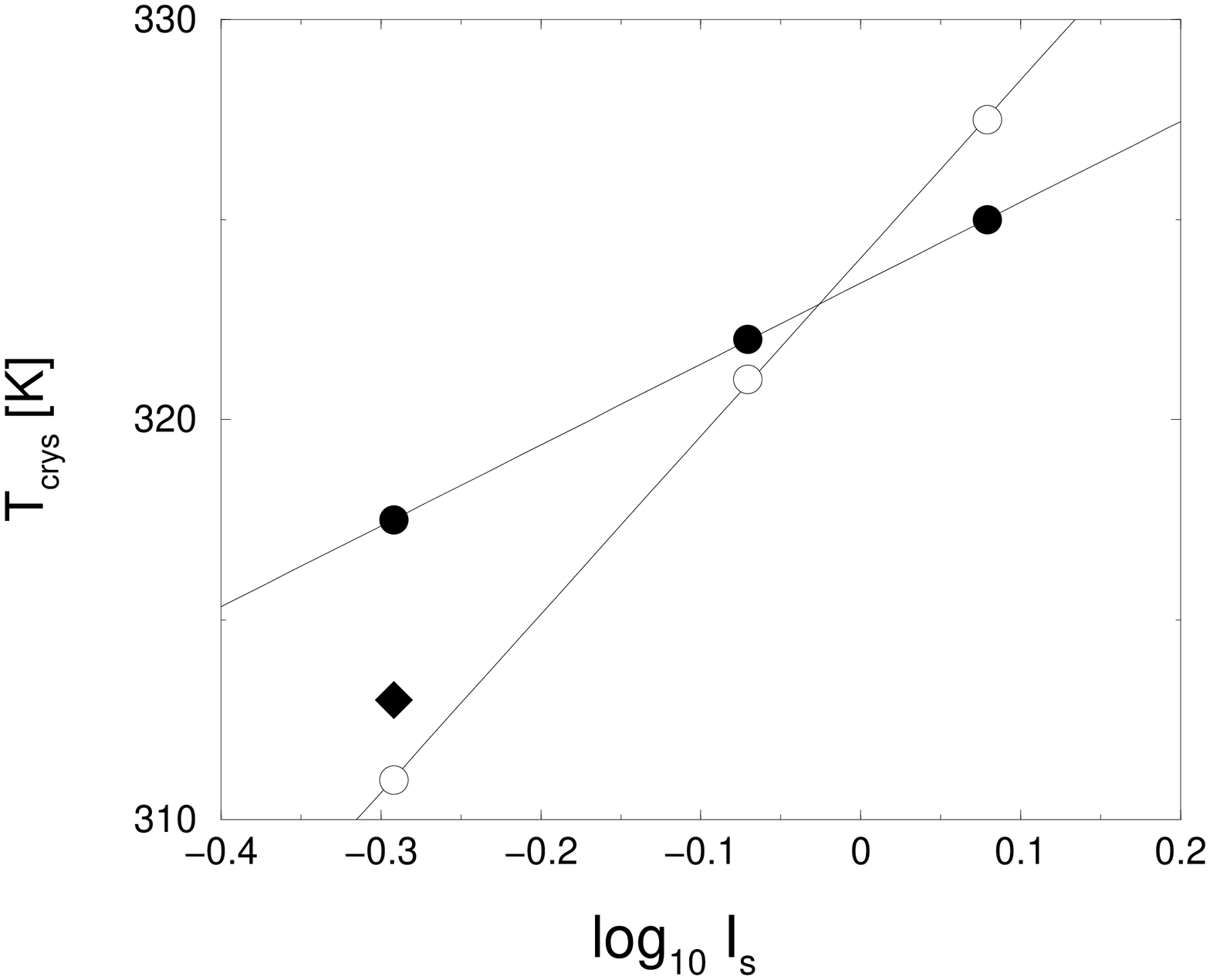}}}
\caption{Critical (left) and solubility (right)
points as a function of the
ionic strength for lysozyme solutions
at fixed protein concentration $\rho=90$~g/l.
Full symbols: experimental data from reference~\cite{muschol97}
(squares),~\cite{broide}
(circles, pH=7.8), and~\cite{muschol97} (diamond, pH=4.5).
Open symbols: DLVO results.
Curves are linear fits of theoretical and experimental data.
}\label{fig:lyso3}
\end{center}
\end{figure}

As visible from figure~\ref{fig:lyso3},
  satisfactory results
are also obtained for
the variation of cloud and solubility temperatures
as a function of the ionic strength.
These results
are reported in a very recent work~\cite{noi4b}
where we have fitted the Hamaker constant
against the collective diffusion coefficient measurements carried out
by
Beretta and coworkers~\cite{beretta}.
It appears that the theory substantially capture the
experimental dependence~\cite{muschol97,broide} of the critical and solubility
temperatures on the logarithm of the ionic strength.

In summary, although a protein
model where we only consider central forces
might appear
oversimplified,
a reasonable prediction of the phase diagram of the
lysozyme solution
may be obtained through a ``colloidal'' model, provided  a judicious fit
of the experimental
second virial coefficient
of the solution~\cite{zamora,muschol95,sic} is performed.

On the other hand, it is evident that a detailed analysis of
the complex behaviour of protein solutions would require a
refined, highly specific
treatment of the microscopic interactions; several improvements
to the DLVO-like model have been recently proposed as,
 for instance,  different  shape~\cite{shape}
or multi-site representations of protein interactions (see e.g.
references~\cite{sear,aelotopic,lowen,sandler}). In comparison
with such studies, our results support the observation that in the
fluid phase, close to the critical point, a distributed approach,
as envisaged for instance in~\cite{aelotopic} may be well
approximated by its spherically averaged counterpart.

\section{Conclusions}

  We have reported results of extensive
investigations of short-range potential models suited to
describe, in a realistic albeit qualitative manner, the physical
properties of prototype globular protein solutions
like for instance lysozyme in water and added salt.
We have focused on the phase diagram of these systems,
and first reviewed predictions
from simulations and integral equation theories of the fluid state
for the hard-core Yukawa fluid.
 We have carried out
a systematic assessment of various refined theories~---~like
the MHNC, the SCOZA
and the GMSA~---~against simulation data, and investigated
the improvement of the strategy adopted to enforce
their thermodynamic consistency.
 It emerges that
integral equations
can reach a good degree of accuracy in the prediction of
phase coexistence properties.
  In particular, 
the basic property of the phase diagram in
many protein solutions,  namely
 a liquid-vapour coexistence line metastable
with respect to the fluid-solid equilibrium,
is correctly reproduced.
Hopefully,
these theoretical approaches may be
used as a complementary tool, especially when
simulations are applied with difficulty, as
in the treatment of
multicomponent systems characterised by the high dilution of one of
the components, or by strong charge and/or size asymmetries.

  We have also reviewed a recent molecular dynamics investigations
of crystallization processes
in a generalised Lennard-Jones model of
protein solutions. This study shows that some
features of the true crystal formation kinetics
are qalitatively represented,
even if
a realistic description of the
natural process (especially as far as the crystallization time scale
is concerned) is still out of
the possibilities offered by such an approach.

   We have then reported on an investigation of protein solutions
based on the DLVO  potential, widely adopted in the
representation of colloidal systems.
 We have shown that
the phase coexistence lines of lysozyme solutions
in water and sodium chloride added salt
can be reproduced through a combined use of theories and
simulations, when the DLVO potential
is parameterised on the experimental
second virial coefficient of the
protein solution at moderate ionic strengths.
Preliminary results
for $\gamma$-crystallin in water and added sodium
phosphate also turn out quantitatively accurate.
  Our procedure is therefore potentially suited for a variety
of protein systems, and presently its extensive 
test appears to us
as essentially hampered by the scarcity of phase
diagram and $B_2$ measurements in the same solution conditions.

In conclusion, basic assumptions on the effective
interaction among globular proteins in solution
are sufficient to accurately reproduce the phase
boundaries of such systems.  In particular,
a parameterisation of a central, pair potential,
based on the experimental second virial coefficient
allows for a confident prediction of the true cloud-points and
solubility lines.
Since a much less effort is required for the experimental
determination of $B_2$,
the advantage of the procedure here exploited
becomes evident.
Our predictions
are accurate also for high protein concentrations,
although experimental measurements of $B_2$
are carried out in
the dilute regime. These findings agree with the
linearity observed in several Debye plots of scattering data~\cite{muschol95},
indicating that $B_2$ values measured in under-saturated
solutions slightly change in the super-saturated regime.

\appendix
\section*{Appendix: Theoretical methods}
\setcounter{section}{1}

The properties of model fluids discussed
in this work have been determined theoretically
in the context of integral
equation theories of liquids~\cite{hansen} like
the modified hypernetted chain
approximation (MHNC,~\cite{MHNC}), the
generalised mean spherical approximation 
(GMSA,~\cite{GMSA}), the
self consistent
Ornstein-Zernike approximation (SCOZA,~\cite{pini}),
and the Hybrid Mean
Spherical Approximation (HMSA,~\cite{HMSA}).
In the MHNC scheme one considers
the Ornstein-Zernike equation
\be\label{hij}
h(r) = c(r) + \rho
\int c(|{\bf r}-{\bf r'}|) h (r') d {\bf r'} \ ,
\ee
(where $\rho$ is the density of the fluid,
$g(r)$ is the radial
distribution function, rdf, $h(r)=g(r)-1$ and $c(r)$
are the pair and direct correlation function, respectively)
coupled to the formally exact expansion result for the rdf
\be\label{gij}
g(r) = \exp [-\beta v(r) + h(r) - c(r) + B(r)] \,
\ee
where $v(r)$ and $B(r)$ are the pair potential and
the bridge function, respectively, of the system under
study, while $\beta= 1/k_{\rm B}T$ with
$k_{\rm B}$ the Boltzmann constant and
 $T$ the temperature.
A closure to equation~(\ref{hij}) is then obtained through 
equation~(\ref{gij}) by
approximating the bridge function for the potential $v(r)$ in terms 
of that of a hard sphere fluid characterized by a hard core diameter
$\sigma^*$ chosen so as to enforce the (partial)
thermodynamic consistency of the theory. 
 
In the GMSA, a solution to the Ornstein-Zernike equation
(2) is obtained by hinging on the exact core condition $g(r)=0$ 
for $r < \sigma$, descending from the fact that $ v(r)= \infty$ on such
a distance range, coupled to the approximation
for the direct correlation function outside the hard-core diameter:
\be \label{cij}
c(r) = -\beta v(r) + K \exp[-z(r - \sigma)]/r \qquad r \ge \sigma
\ee
where $K$ and $z$ are used as adjustable parameters in order to
impose the internal thermodynamic consistency of the theory.

In the SCOZA scheme $c(r)$ is written as:
\be \label{8} 
c(r) = - A\beta v(r) +K_{\rm HS}\exp [-z_{\rm
HS}(r-\sigma)]/r \qquad r \ge\sigma \,, \ee
$K_{\rm HS}$ and $z_{\rm HS}$ are
predetermined by setting $v(r)=0$ in equation~(\ref{8}) 
and requiring that both
the compressibility and the virial routes to thermodynamics yield the
Carnahan-Starling equation of state for a hard-sphere fluid.  
The factor $A$ is then obtained by imposing that the compressibility and
energy routes yield the same thermodynamics.

The HMSA is so called because 
it interpolates between
the hypernetted chain (HNC)~\cite{report} approximation and the
soft mean spherical approximation (SMSA)~\cite{report}:
it is written as
\be\label{hmsa} \fl
g(r) = \exp [-\beta v_{1}(r)] \: \left\{1 + \frac{ \exp \:\{\:f(r)
\: [\: h(r) -c(r)\ -\beta v_{2}(r) ]\:\} - 1] } {f(r)}\right\}
\ee
In equation~(\ref{hmsa}) the potential $v(r)$ is splitted into a soft
core repulsive part, [$v_1(r)$], and an attractive part, [$v_2(r)$],
according to:
\be\label{pothmsa1}  \fl
v_{1}(r) \ =\cases{v(r)-v(r_{\rm min})& 
$r \leq r_{\rm min}$ \cr 0 & $ r > r_{\rm min} $
\cr} \qquad
v_{2}(r) \ =\cases{v(r_{\rm min})& $r \leq r_{\rm min}$ \cr
 v(r) & $ r > r_{\rm min} $ \cr} 
\ee
where $r_{\rm min}$ denotes the position of the minimum of 
potential $v(r)$ and $f(r)=1-\exp[\xi r]$.
The parameter $\xi$ is used to enforce the thermodynamic consistency.

In order to determine theoretically the properties 
of the solid phase 
of the Girifalco \C60 model we have employed
a standard perturbation theory (PT,~\cite{hansen}).
In this scheme, the model's free energy per particle, $f$,
is calculated as a first-order expansion around
the free energy
of a reference hard-sphere solid, $f_{\rm HS}$,
according to:
\be\label{eq:exp}
\beta f=\beta f_{\rm HS}+ \frac{\beta \rho}{2}
\int v_{\rm pert}(r) \ \overline g_{\rm HS}(r) \diff {\mathbf r}
\,, \ee
where $\overline g_{\rm HS}(r)$ is the radial
distribution function of the reference hard-sphere solid.
In order to determine the perturbative part
of the potential, $v_{\rm pert}(r)$ and
the diameter $\sigma^*$ of the reference hard-sphere system
entering equation~(\ref{eq:exp}), we have followed 
the WCA prescription~\cite{chandler},
that amounts to 
split the original potential according to 
equation~(\ref{pothmsa1}) with $v_{\rm pert}(r) \equiv v_2(r)$
and $v_{\rm ref} \equiv v_1(r)$; then~\cite{BH}:
\be\label{eq:sigma1}
\sigma^*=\int_0^\infty \{1 -\exp[-\beta v_{\rm ref}(r)]\}\rmd r \,.
\ee

\end{document}